\documentclass[twocolumn,prl,aps,superscriptaddress,altaffilletter]{revtex4}
\usepackage{graphicx}
\usepackage{epsfig}
\usepackage{amsfonts,amsmath}

\def\fsec    {fs}                               

\def\mvC     {{\mbox{$\mathrm{MeV/c^2}$}}}
\def\mvc     {{\mbox{$\mvC \ $}}}

\def\dsj     {{\mbox{$\mathrm{D^{+}_{sJ}}$}}}
\def\dsM     {{\mbox{$\rm{D_{s}}$}}}
\def\dsm     {{\mbox{$\dsM\ $}}}

\def\dS      {{\mbox{$\rm{K}\rm{K}\pi^{\pm} $}}}
\def\ds      {{\mbox{$\dS \ $}}}
\def\dse     {{\mbox{$\dsj(2632) \rightarrow \rm{D_{s}^{+}} \eta \ $}}}
\def\dzk     {{\mbox{$\rm{D^{0} K^+} $}}}
\def\dseta   {{\mbox{$\rm{D_{s}^{+} \eta} \ $}}}

\def\Signfe  { $13.3 \ \sigma$ }                
\def\Nsige   {5087 }                            
\def\dNsige  {863 }                             
\def\Masse   {544.8 }                           
\def\sMasse  {2.9 }                             
\def\Rese    {28 }                              
\def\sRese   {4 }                               
\def\Rsime   {30 }                              
\def\Rchie   {1.17 }                            

\def\Nevt    {101 }                             
\def\Back    {54.9 }                            
\def\dBack   { 1.5 }                            
\def\Esig    { 46.1 }                           
\def\Signf   {$6.2 \ \sigma$ }                  
\def\Nsig    {43.4 }                            
\def\dNsig   {9.1 }                             
\def\DMass   {666.9 }                           
\def\Mass    {2635.4 }                          
\def\sMass   {3.3 }                             
\def\Res     {10.9 }                            
\def\Rchi    {1.10 }                            

\def\Signfa  {$5.4 \ \sigma$ }                  
\def\Nsiga   {25 }                              
\def\dNsiga  { 9 }                              
\def\DMassa  {705.4}                            
\def\Massa   {2569.9 }                          
\def\Widtha  {$14^{+9}_{-6}$}                   
\def\sMassa  {4.3}                              
\def\Resa    {4.9 }                             
\def\Rchia   {0.77 }                            

\def\Nevtb   {21 }                              
\def\Backb   {6.9 }                             
\def\Signfb  {$5.3 \ \sigma$ }                  
\def\Nsigb   {13.2}                             
\def\dNsigb  { 4.9}                             
\def\DMassb  {767.0 }                           
\def\Massb   {2631.5 }                          
\def\sMassb  {2.0 }                             
\def\Widthb  {<17}                              
\def\Resb    {4.9 }                             

\def\Signfd  {$4.8 \ \sigma$ }                  
\def\Nsigd   {60.1}                             
\def\dNsigd  {14.9}                             
\def\DMassd  {143.7 }                           
\def\Massd   {2112.0 }                          
\def\sMassd  {1.5 }                             
\def\Resd    {5.9 }                             
\def\Rchid   {0.94 }                            

\def\MassA   {2632.5 }                          
\def\sMassA  {1.7 }                             

\def\Br      {0.14}                             
\def\dBr     {0.06}                             

\begin{document}
\title{First Observation of a Narrow Charm-Strange Meson \dse and 
       $\rm{D^0 K^+}$}


\affiliation{Ball State University, Muncie, IN 47306, U.S.A.}
\affiliation{Bogazici University, Bebek 80815 Istanbul, Turkey}
\affiliation{Carnegie-Mellon University, Pittsburgh, PA 15213, U.S.A.}
\affiliation{Centro Brasileiro de Pesquisas F\'{\i}sicas, Rio de Janeiro, Brazil}
\affiliation{Fermi National Accelerator Laboratory, Batavia, IL 60510, U.S.A.}
\affiliation{Institute for High Energy Physics, Protvino, Russia}
\affiliation{Institute of High Energy Physics, Beijing, P.R. China}
\affiliation{Institute of Theoretical and Experimental Physics, Moscow, Russia}
\affiliation{Max-Planck-Institut f\"ur Kernphysik, 69117 Heidelberg, Germany}
\affiliation{Moscow State University, Moscow, Russia}
\affiliation{Petersburg Nuclear Physics Institute, St. Petersburg, Russia}
\affiliation{Tel Aviv University, 69978 Ramat Aviv, Israel}
\affiliation{Universidad Aut\'onoma de San Luis Potos\'{\i}, San Luis Potos\'{\i}, Mexico}
\affiliation{Universidade Federal da Para\'{\i}ba, Para\'{\i}ba, Brazil}
\affiliation{University of Bristol, Bristol BS8~1TL, United Kingdom}
\affiliation{University of Iowa, Iowa City, IA 52242, U.S.A.}
\affiliation{University of Michigan-Flint, Flint, MI 48502, U.S.A.}
\affiliation{University of Rome ``La Sapienza'' and INFN, Rome, Italy}
\affiliation{University of S\~ao Paulo, S\~ao Paulo, Brazil}
\affiliation{University of Trieste and INFN, Trieste, Italy}
\author{A.V.~Evdokimov}
\affiliation{Institute of Theoretical and Experimental Physics, Moscow, Russia}
\author{U.~Akgun}
\affiliation{University of Iowa, Iowa City, IA 52242, U.S.A.}
\author{G.~Alkhazov}
\affiliation{Petersburg Nuclear Physics Institute, St. Petersburg, Russia}
\author{J.~Amaro-Reyes}
\affiliation{Universidad Aut\'onoma de San Luis Potos\'{\i}, San Luis Potos\'{\i}, Mexico}
\author{A.G.~Atamantchouk}
\altaffiliation{Deceased}
\affiliation{Petersburg Nuclear Physics Institute, St. Petersburg, Russia}
\author{A.S.~Ayan}
\affiliation{University of Iowa, Iowa City, IA 52242, U.S.A.}
\author{M.Y.~Balatz}
\altaffiliation{Deceased}
\affiliation{Institute of Theoretical and Experimental Physics, Moscow, Russia}
\author{N.F.~Bondar}
\affiliation{Petersburg Nuclear Physics Institute, St. Petersburg, Russia}
\author{P.S.~Cooper}
\affiliation{Fermi National Accelerator Laboratory, Batavia, IL 60510, U.S.A.}
\author{L.J.~Dauwe}
\affiliation{University of Michigan-Flint, Flint, MI 48502, U.S.A.}
% ...
\collaboration{The SELEX Collaboration}
\noaffiliation

\begin{abstract}
We report the first observation of a charm-strange meson 
$\dsj(2632)$ at a mass of $\MassA \pm \sMassA\ \mvc$ in data from SELEX, 
the charm hadro-production experiment E781 at Fermilab.  This state is seen in 
two decay modes, \dseta and \dzk.  In the $\rm{D_{s}^{+} \eta}$ decay mode 
we observe a peak with \Nevt events over a combinatoric background of \Back 
events at a mass of \Mass $\pm$ \sMass \mvC.  There is a corresponding peak of 
\Nevtb events over a background of \Backb at \Massb $\pm$ \sMassb \mvc in the 
decay mode \dzk.  The decay width of this state is $\Widthb$ \mvc at $90\%$ 
confidence level.  The relative branching ratio $\Gamma(\dzk)/\Gamma(\dseta)$ 
is \Br \ $\pm$ \dBr.  The mechanism which keeps this state narrow is unclear.  
Its decay pattern is also unusual, being dominated by the \dseta decay mode. 

\end{abstract}

\maketitle

%
\newpage
\newpage

In 2003 the BaBar collaboration reported the first observation of a massive,
narrow charm-strange meson \dsj(2317) below the $\rm{D K}$ 
threshold~\cite{Babar}.  Confirmation quickly followed from 
CLEO~\cite{CLEO} and BELLE~\cite{Belle}.
The CLEO collaboration showed that a higher-lying state, 
suggested by BaBar, existed and was a partner to the 
\dsj(2317).  A number of theory papers suggested different explanations for
the unexpectedly low mass of the state, which had been thought to lie above
the DK threshold~\cite{nowak,nowak1,nowak2,estia,hill,hill1}.  A prediction 
relevant to this experiment was that the pattern of parity-doubled states 
is expected to continue to higher excitations with similar 
splittings~\cite{Bill}.  

%

The SELEX experiment used the Fermilab charged hyperon beam at 
\mbox{600\,GeV/c} to produce charmed particles in a set of thin foil targets 
of Cu or diamond.  The negative beam composition was approximately half 
$\Sigma^-$ and half $\pi^-$.  
The three-stage magnetic spectrometer is shown elsewhere~\cite{Thesis,SELEX}. 
The most important features are the high-precision, highly redundant, vertex 
detector that provides an average proper time resolution of \mbox{20\,\fsec} 
for charm decays, a \mbox{10\,m} long Ring-Imaging Cerenkov (RICH) 
detector that separates $\pi$ from K up to \mbox{165\,GeV/c}~\cite{RICH}, and 
a high-resolution tracking system that has momentum resolution of 
\mbox{$\sigma_{p}/p$\,$<1\%$} for a \mbox{150\,GeV/c} track.  Photons are 
detected in 3 lead glass photon detectors, one following each spectrometer 
magnet.  The photon angular coverage in the center of mass 
typically exceeds 2$\pi$.  For this analysis the photon energy threshold was 
\mbox{2\,GeV}.  Previous SELEX $\rm{D_s}$ studies
showed that most of the signal came from the $\Sigma^-$ beam~\cite{prod}.
We restrict ourselves in this analysis to the $10 \times 10^{9}$ 
$\Sigma^{-}$-induced interactions.

%

In this study we began with the SELEX 
$\rm{D_s^{\pm} \rightarrow \rm{K}^+ \rm{K}^-} \pi^{\pm}$ sample used in 
lifetime and hadroproduction studies~\cite{lifetime,prod}. 
Charged conjugate final states are included here and throughout this paper.  
The sample-defining cuts are defined in the references.  
The \dsm meson momentum vector had to point back to the primary vertex with 
\mbox{$\chi^2$\,$<8$} and its decay point must have a vertex separation 
significance of at least 8 from the primary.   Tracks which traversed the RICH
($p\,{\scriptstyle\gtrsim}\,22\,\mathrm{GeV}/c$) were identified as
kaons if this hypothesis was most likely.  The pion was 
required to be RICH-identified 
if it went into its acceptance.  
There are $544 \pm 29$ $\Sigma^-$ induced signal events with these cuts.

Due to high multiplicity, photon detection in an open charm trigger is 
challenging.  SELEX has 3 lead glass calorimeters covering much of the 
forward solid angle.  The energy scale for the detector 
was set first by using electron beam scans.  Then $\pi^{0}$ decays were 
reconstructed from exclusive trigger data, which selected low-multiplicity 
radiative final states:
$\eta \rightarrow \gamma \gamma \ \rm{and} \ \pi^+ \pi^- \pi^0, \  \omega 
\rightarrow \pi^+ \pi^- \pi^0$ as well as $\eta^{\prime}$ and f(1285) 
mesons~\cite{TM}.  The final energy scale corrections were developed using 
$\pi^{0}$ decays from the high-multiplicity charm trigger data. Further 
checks in the charm data set were made using single
photon decays, e.g., $\Sigma^0 \rightarrow \Lambda \gamma$. The uncertainty 
in the photon energy scale is less than 2\%.  Details can be found in 
Ref.~\cite{TM}.  

\begin{figure}[ht]
\includegraphics[width=2.75in]{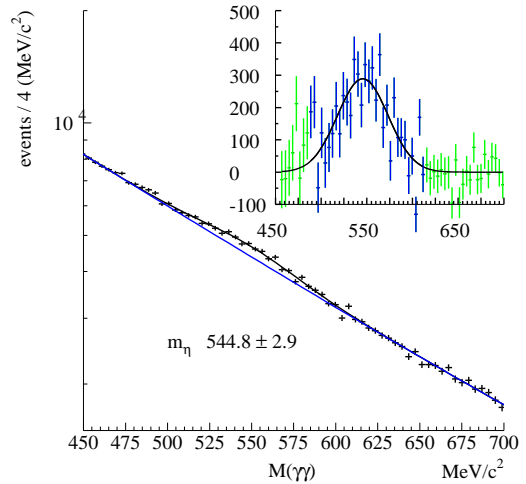}
\caption{M($\gamma\gamma$) distribution for photon pairs in the $\eta$ mass 
         region from 0.1\% of the data sample. Results for the fit shown 
         are in Table~\ref{fits}.  The inset shows the background subtracted 
         $\eta$ signal.   The dark points indicate the $\eta$ signal region. } 
\label{eta}
\end{figure}

We selected $\eta \rightarrow \gamma\gamma$ candidates in the $\gamma 
\gamma$ mass range 400-800 \mvC.  Each photon of the pair has 
\mbox{$E_{\gamma}$\,$>$2 GeV}. The photon pair has 
\mbox{$\rm{E}_{\gamma\gamma}$\,$> $15 GeV}.  The $\gamma\gamma$ mass 
distribution from $10^6$ charm-trigger events (0.1\% of the data) is shown in 
Fig.~\ref{eta} where a $\eta$ signal over a large combinatoric background 
is seen.  A fit to a Gaussian plus an exponentially falling background yields 
an $\eta$ mass of \Masse $\pm$ \sMasse~\mvc consistent with the PDG 
value~\cite{PDG}.  The mass uncertainty for this and all subsequent states 
is only statistical.

The observed resolution is $\Rese \pm \sRese $~\mvc consistent with the SELEX 
simulation result, \Rsime \mvC.  The simulation includes all the material in 
the spectrometer and also reproduces the observed $\pi^{0}$ width as a 
function of energy~\cite{TM}.  
The $\eta$ signal fraction is $5\%$ within a $\pm60 \mvc$ mass window when 
we eliminate events with more than 5 $\eta$ candidates in this mass window.

\begin{table*}[t]
\centering
\begin{ruledtabular}
\begin{tabular}{|c|c|c|c|c|c|c|c|c|}
Fig.& state
    & fitted
    & $\Delta$M
    & Mass
    & Significance
    & $\sigma$
    & $\Gamma$
    & $\chi^{2}/n_{d}$\\

    & 
    & yield
    & \mvc
    & \mvc
    & $(S-B)/\surd{\overline{B}}$
    & \mvc
    & \mvc
    & \\
\hline

 1&   $\eta(548)   \rightarrow \gamma\gamma$
  &   $\Nsige \pm \dNsige$
  &                     
  &   $\Masse \pm \sMasse$
  &   \Signfe
  &   $\Rese  \pm \sRese$
  &
  &   \Rchie\\

 2& $D_{s}^{+}(2632) \rightarrow D_s^+ \eta$
  &   $\Nsig  \pm \dNsig$
  &   $\DMass \pm \sMass$
  &   $\Mass  \pm \sMass$
  &    \Signf
  &   $\Res  $
  &
  &   \Rchi\\

 3& $D_{s}^{+}(2573) \rightarrow D^0 K^+$
  &   $\Nsiga  \pm  \dNsiga$
  &   $\DMassa \pm \sMassa$
  &   $\Massa \pm \sMassa$
  &   \Signfa
  &   \Resa
  &   \Widtha
  &   \Rchia\\

 3& $D_{s}^{+}(2632) \rightarrow D^0 K^+$
  &   $\Nsigb  \pm  \dNsigb$
  &   $\DMassb \pm  \sMassb$
  &   $\Massb  \pm  \sMassb$
  &   \Signfb
  &   $\Resb$
  &   $\Widthb (90\%CL)$
  &   \\

  & $D_{s}^{*+}(2112) \rightarrow D_s^+ \gamma$
  &   $\Nsigd  \pm \dNsigd$
  &   $\DMassd \pm \sMassd$
  &   $\Massd  \pm \sMassd$
  &    \Signfd
  &   $\Resd  $
  &
  &   \Rchid\\

\end{tabular}
\end{ruledtabular}
\caption{Fit results for Figures 1-3 and 
         $D_{s}^{*+}(2112) \rightarrow D_s^+ \gamma$ .}
\label{fits}
\end{table*}

We searched for high-mass charm-strange decays that followed the pattern 
\dsm plus pseudoscalar meson.  We had good acceptance and efficiency for the 
\dsm $\eta$ channel.  The event selection used included the eta selection 
above, and the $D_s$ selection described above, which yields a S/N of 4/1.
The \dsm momenta are typically \mbox{150\,GeV/c} in the SELEX data set; 
the $\rm{E}_{\eta} >$ 15 GeV energy cut is very loose.  We rejected 
events in which there were more than 5 $\eta$ candidates in the signal region. 
This cut removed 18 \dsm candidates (3.3\%) while reducing the $\eta$ 
candidate list by 20\%.  The $\eta$ signal region is shown in Fig.~\ref{eta}.  
The final sample consisted of 615 $\eta$ candidates from 526 \dsm candidates.

\begin{figure}[ht]
\includegraphics[width=2.75in]{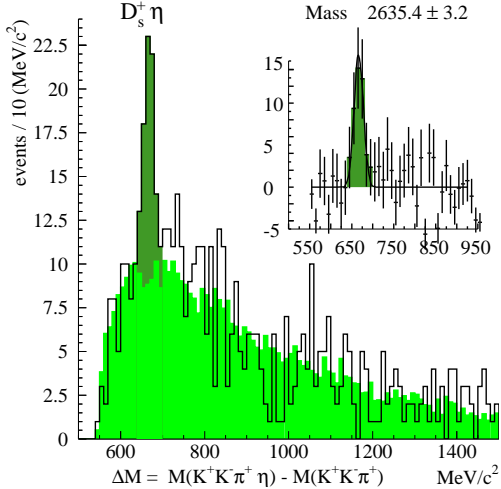}
\caption{ M(\ds $\eta$) - M(\ds) mass difference distribution.  
          Charged conjugates are included.
          The darker shaded region is the event excess used in the 
          estimation of  signal significance. The lighter shaded region is 
          the event-mixed combinatoric background described in the text.  
          The inset shows the difference of the two with a Gaussian fit to 
          the signal.  Results for the fit shown are in Table~\ref{fits}. }
\label{dseta}
\end{figure}

The results of our search are shown in the M(\ds $\eta$) - M(\dS)
mass difference distribution in Fig.~\ref{dseta}(a).  In this plot we 
fixed the $\eta$ mass at the PDG value~\cite{PDG} by 
defining an $\eta$ 4-vector with the measured $\eta$ momentum and the 
PDG $\eta$ mass.   A clear peak is seen at a mass difference of 
\DMass $\pm$ \sMass \mvC. 

To estimate the combinatoric background, we matched each \dsm candidate with 
$\eta$ candidates from 25 other sample events to form a event-mixed sample 
representing the combinatoric background of true single charm production and 
real $\eta$ candidates.  The event-mixed mass distribution was then scaled 
down by 1/25 to predict the combinatoric background in the signal channel.  
As can be seen in Fig.~\ref{dseta}, the event-mixed background models the 
background shape very well, but produces no signal peak.  

To estimate the signal yield we subtracted the combinatoric background 
(light shaded area) from the signal data.  The resulting difference histogram 
is plotted in the inset in Fig.~\ref{dseta} in the mass-difference range 
appropriate to our search (\dsj \ masses up to 2900 \mvC).  Outside the peak 
region the data scatter about 0.   The width in the $\rm{D^0 K^+}$ mode, 
to be discussed below, is consistent with a \Resb \mvc Gaussian.  We are 
insensitive to the natural width in the \dseta mode.  Therefore we fit the 
difference histogram with only a Gaussian with no residual background terms.
The Gaussian width is fixed at the simulation value of \Res \mvC.  The fit 
yield is \Nsig $\pm$ \dNsig events at a mass of \Mass $\pm$ \sMass \mvC.  The 
reduced $\chi^2$ is \Rchi with a confidence level of 31\%.  

To assess the Poisson fluctuation probability to observe this excess we note 
that the resolution is about 1 bin; we take a 6-bin cluster as a signal region
and perform a counting experiment.  There are \Nevt events over a background of 
\Back$ \pm $\dBack events giving an excess of \Esig events in 6 adjacent bins.
The Poisson fluctuation probability for this excess is $3 \times 10^{-7}$ 
including the uncertainty in the background.  A conservative estimate of the 
fluctuation probability anywhere in the search region (up to 2900 \mvC) is 
$6 \times 10^{-6}$.

The signal does not change with  variations of $\pm \ 2\%$ in the photon 
energy scale. We also studied combinations of events in the $D_s$ mass 
sidebands with $\eta$ candidates and candidates in the $D_s$ mass peak 
with events in the $\eta$ mass sidebands.  In all cases only smooth 
combinatoric backgrounds, as in Fig.~\ref{dseta}, were observed.  The 
$\eta$ population in this sample, corrected for $\eta$'s from the 
signal channel, has a signal fraction of 0.12 $\pm$ 0.05, 1.4$\sigma$ higher 
than the global average of 0.05.  It is not possible to separate statistical 
fluctuation from a possible $\eta$ enrichment of the overall \dsm sample.

%

A GEANT simulation was also used to determine the overall acceptance
for these signals.  If we detected the \dsm from a \dse decay, then
35 $\pm$ 2\% of the time we also detected the 
$\eta \rightarrow \gamma \gamma$.  This acceptance was obtained by 
embedding \dse decay events in existing events from the real data.  About 
55\% of the \dsm decays in SELEX come through this high mass state.
For comparison, about 25\% of the \dsm come from \dsm(2112) decays
and a small fraction from either \dsj(2317) or \dsj(2460) decays,
which are marginally visible in SELEX data.

The decay \dsj(2632) $\rightarrow \dzk$ is kinematically allowed.  
After finding the \dse signal we searched for this second decay mode as 
confirmation.  The $D^0$ sample is the $\Sigma^-$ induced 
$D^0 \rightarrow K^{-}\pi^{+}$  subset of the sample used in our measurement 
of the $D^0$ lifetime~\cite{d0} with tight $D^0$ cuts 
($L/\sigma>6$, point-back $\chi^2<5$, and a good fit to the secondary vertex;
$\chi^2/n_d<3$).  The $K^+$ track is $>$ 46 GeV/c (RICH kaon threshold) and is 
strongly identified by the RICH as \mbox{$\ge$\,10} times more likely to be a 
kaon than any other hypothesis. 

\begin{figure}[ht]
\includegraphics[width=2.75in]{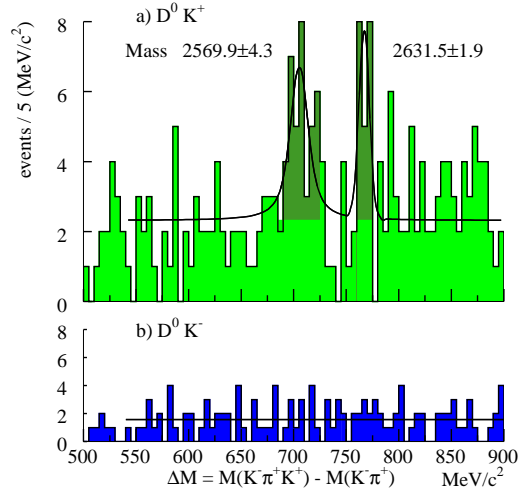}
\caption{(a) \dsm(2632) $\rightarrow \dzk$ mass difference 
             distribution. 
             Charged conjugates are included. 
             The shaded regions are the event excesses used in the estimation 
             of  signal significances. 
             Results for the fit shown are in Table~\ref{fits}.
        (b) Wrong sign background $\rm{D^0 \ K^{-}}$ events, as described in 
             the text.}
\label{d0k}
\end{figure}

The results are shown in Fig.~\ref{d0k}(a) where we see both the known 
\dsj(2573) state clearly and another peak above the \dsj(2573).  
We fit each peak with a Breit-Wigner convolved with a fixed width Gaussian 
plus a constant background term (as suggested from the wrong-sign data
discussed below).  The Gaussian resolution is set to the simulation 
value of $\Resb$~\mvC.  The mass difference and width of the \dsj(2573) 
returned by the fit, $\Delta M$ = \DMassa $\pm$ \sMassa~\mvc and 
$\Gamma$ = \Widtha~\mvc, respectively, agree well with the 
PDG values~\cite{PDG} of $\Delta M$ = 707.9 $\pm$ 1.5~\mvc and 
$\Gamma = 15^{+5}_{-4} \mvC$.  The fitted mass difference of the second 
Breit-Wigner is \DMassb $\pm$ \sMassb \mvC, leading to a mass for the new 
peak of \Massb $\pm$ \sMassb \mvC.  
The fitted yield is $\Nsigb \pm \dNsigb $ events.  The signal spread is 
consistent with the Gaussian resolution, even when plotted in 2.5 \mvc bins,
limiting the possible natural width.
For the Breit-Wigner fit we find a limit for the width of $\Widthb$ \mvc at 
$90\%$ confidence level.  This signal has a significance 
$([S-B]/\surd{\overline{B}})$ of \Signfb in a $\pm 15 \ \mvc$ interval.  The 
mass difference between this signal and the one seen in the \dseta mode is 
3.9 $\pm$ 3.8 \mvc, statistically consistent with being the same mass.  
Unlike the $\rm{D_s}$ case, the $\rm{D^0 K^+}$ decay contributes a small
fraction to the SELEX $\rm{D^0 }$ sample. 

Combinatoric background will be equally likely to produce a $\rm{D^0 K^-}$
combination (wrong-sign kaon) as a \dzk.  The wrong-sign combinations are 
shown in Fig.~\ref{d0k}(b).  There is no structure in these data, which fits
well to a constant background.  We conclude 
that the peak at \Massb \mvc is real and confirms the observation in the 
\dseta mode.  

The relative branching ratio $\Gamma(\dzk)/\Gamma(\dseta)$ must be corrected 
for the relative acceptances of $D^0$, $D_s$, $\eta$, and $K^+$ mesons, 
for the $\eta$, $D^0$ and $D^{+}_{s}$ branching ratios, the relative 
acceptances of the \dsj(2632) final states.  We estimate the relative 
acceptance ratio from simulation as $91 \pm 3\%$.  The relative branching 
ratio is \quad $\Gamma(\dzk)/\Gamma(\dseta)$ = \Br \ $\pm$ \dBr.  \quad
Relative phase space favors the \dzk mode by a factor of 1.53, making this
low branching ratio even more surprising.

%

In conclusion we combined our clean sample of \dsm mesons with additional 
photon pairs which made $\eta$ candidates to study the $\rm{D_s} \eta$
mass spectrum.  We observe a clear peak of \Nsig $\pm$ \dNsig events with a 
Poisson fluctuation probability $<$ 6 x $10^{-6}$ at a mass difference 
of \DMass $\pm$ \sMass~\mvc above the ground state \dsM.  The background shape 
is well-represented by combinatoric background from event-mixing, as discussed 
above.   A corresponding mass peak is also seen in the \dzk channel 
with a significance of \Signfb at the same mass.  The combined measurement of 
the mass of this state is \MassA $\pm$ \sMassA \mvC.  The SELEX mass scale
systematic error is under study, but all charm meson masses measured in
SELEX agree with PDG values within 1 \mvC.  The state is very
narrow, consistent with a width due only to resolution in the \dzk decay mode. 
The 90\% confidence level upper limit for the width is $\Widthb$~\mvC.

SELEX reports these peaks as the first observation of yet another narrow, 
high-mass \dsm state decaying strongly to a ground state charm plus a 
pseudoscalar meson.  The mechanism which keeps this state narrow is unclear. 
The $\rm{D^0 K^+}$ channel is well above threshold, with a Q value 
$\sim$~275 MeV.  The branching ratios for this state are also unusual.  
The \dseta decay rate dominates the \dzk rate by a factor of $\sim$~7 despite 
having half the phase space.  

%
The authors are indebted to the staff of Fermi National Accelerator Laboratory
and for invaluable technical support from the staffs of collaborating
institutions.  We thank A.M.~Badalyan, Bill Bardeen, Estia Eichten, 
Chris Hill, Maciej Nowak and Yu.A.~Simonov for helpful conversations.
We gratefully acknowledge helpful suggestions, comments and criticisms from a 
number of members of the charm community.
This project was supported in part by Bundesministerium f\"ur Bildung, 
Wissenschaft, Forschung und Technologie, Consejo Nacional de 
Ciencia y Tecnolog\'{\i}a {\nobreak (CONACyT)},
Conselho Nacional de Desenvolvimento Cient\'{\i}fico e Tecnol\'ogico,
Fondo de Apoyo a la Investigaci\'on (UASLP),
Funda\c{c}\~ao de Amparo \`a Pesquisa do Estado de S\~ao Paulo (FAPESP),
the Israel Science Foundation founded by the Israel Academy of Sciences and 
Humanities, Istituto Nazionale di Fisica Nucleare (INFN),
the International Science Foundation (ISF),
the National Science Foundation (Phy \#9602178),a
NATO (grant CR6.941058-1360/94),
the Russian Academy of Science,
the Russian Ministry of Science and Technology,
the Turkish Scientific and Technological Research Board (T\"{U}B\.ITAK),
the U.S. Department of Energy (DOE grant DE-FG02-91ER40664 and DOE contract
number DE-AC02-76CHO3000), and
the U.S.-Israel Binational Science Foundation (BSF).

{\it Note Added in Proof - }
At the summer conferences the B-factory experiments\cite{Galik,Babar2} and the 
FOCUS photoproduction experiment\cite{Focus} all reported negative results in 
their search for $D_{sJ}$(2632) production.  The SELEX data have an 
unusually low ratio of $R_s \equiv D_s^{*+}(2112)/D_s^+$ production (Table I) 
compared to $e^+e^-$ machines.  The SELEX $R_s$ ratio can be brought into 
agreement with the published CLEO result~\cite{CLEO} if we 
{\it exclude} the $D_s$ events that come through $D_{sJ}$(2632) decays 
(55\% of the SELEX $D_{s}$ yield).  This fact, together with the strong 
production asymmetry in the SELEX data, can be interpreted as two-component 
production of charm-strange states from the $\Sigma^-$ beam.  One component 
involves normal fragmentation like in photoproduction, the other involves 
a different  mechanism connected with the beam hadron.   To understand this 
production conundrum and to place this new state in the spectroscopy of the 
charm-strange meson system will require careful study from a number of 
experiments in the future.

\bibliography{etaprl}

\begin{thebibliography}{21}
\expandafter\ifx\csname natexlab\endcsname\relax\def\natexlab#1{#1}\fi
\expandafter\ifx\csname bibnamefont\endcsname\relax
  \def\bibnamefont#1{#1}\fi
\expandafter\ifx\csname bibfnamefont\endcsname\relax
  \def\bibfnamefont#1{#1}\fi
\expandafter\ifx\csname citenamefont\endcsname\relax
  \def\citenamefont#1{#1}\fi
\expandafter\ifx\csname url\endcsname\relax
  \def\url#1{\texttt{#1}}\fi
\expandafter\ifx\csname urlprefix\endcsname\relax\def\urlprefix{URL }\fi
\providecommand{\bibinfo}[2]{#2}
\providecommand{\eprint}[2][]{\url{#2}}

\bibitem[{\citenamefont{Aubert et~al.}(2003)}]{Babar}
\bibinfo{author}{\bibfnamefont{B.}~\bibnamefont{Aubert}} \bibnamefont{et~al.}
  (\bibinfo{collaboration}{BABAR}), \bibinfo{journal}{Phys. Rev. Lett.}
  \textbf{\bibinfo{volume}{90}}, \bibinfo{pages}{242001}
  (\bibinfo{year}{2003}), \eprint{hep-ex/0304021}.

\bibitem[{\citenamefont{Besson et~al.}(2003)}]{CLEO}
\bibinfo{author}{\bibfnamefont{D.}~\bibnamefont{Besson}} \bibnamefont{et~al.}
  (\bibinfo{collaboration}{CLEO}), \bibinfo{journal}{Phys. Rev.}
  \textbf{\bibinfo{volume}{D68}}, \bibinfo{pages}{032002}
  (\bibinfo{year}{2003}), \eprint{hep-ex/0305100}.

\bibitem[{\citenamefont{Krokovny et~al.}(2003)}]{Belle}
\bibinfo{author}{\bibfnamefont{P.}~\bibnamefont{Krokovny}} \bibnamefont{et~al.}
  (\bibinfo{collaboration}{Belle}), \bibinfo{journal}{Phys. Rev. Lett.}
  \textbf{\bibinfo{volume}{91}}, \bibinfo{pages}{262002}
  (\bibinfo{year}{2003}), \eprint{hep-ex/0308019}.

\bibitem[{\citenamefont{Nowak and Zahed}(1993)}]{nowak}
\bibinfo{author}{\bibfnamefont{M.~A.} \bibnamefont{Nowak}} \bibnamefont{and}
  \bibinfo{author}{\bibfnamefont{I.}~\bibnamefont{Zahed}},
  \bibinfo{journal}{Phys. Rev.} \textbf{\bibinfo{volume}{D48}},
  \bibinfo{pages}{356} (\bibinfo{year}{1993}).

\bibitem[{\citenamefont{Nowak et~al.}(1993)\citenamefont{Nowak, Rho, and
  Zahed}}]{nowak1}
\bibinfo{author}{\bibfnamefont{M.~A.} \bibnamefont{Nowak}},
  \bibinfo{author}{\bibfnamefont{M.}~\bibnamefont{Rho}}, \bibnamefont{and}
  \bibinfo{author}{\bibfnamefont{I.}~\bibnamefont{Zahed}},
  \bibinfo{journal}{Phys. Rev.} \textbf{\bibinfo{volume}{D48}},
  \bibinfo{pages}{4370} (\bibinfo{year}{1993}), \eprint{hep-ph/9209272}.

\bibitem[{\citenamefont{Nowak et~al.}(2003)\citenamefont{Nowak, Rho, and
  Zahed}}]{nowak2}
\bibinfo{author}{\bibfnamefont{M.~A.} \bibnamefont{Nowak}},
  \bibinfo{author}{\bibfnamefont{M.}~\bibnamefont{Rho}}, \bibnamefont{and}
  \bibinfo{author}{\bibfnamefont{I.}~\bibnamefont{Zahed}}
  (\bibinfo{year}{2003}), \eprint{hep-ph/0307102}.

\bibitem[{\citenamefont{Di~Pierro and Eichten}(2001)}]{estia}
\bibinfo{author}{\bibfnamefont{M.}~\bibnamefont{Di~Pierro}} \bibnamefont{and}
  \bibinfo{author}{\bibfnamefont{E.}~\bibnamefont{Eichten}},
  \bibinfo{journal}{Phys. Rev.} \textbf{\bibinfo{volume}{D64}},
  \bibinfo{pages}{114004} (\bibinfo{year}{2001}), \eprint{hep-ph/0104208}.

\bibitem[{\citenamefont{Bardeen et~al.}(2003)\citenamefont{Bardeen, Eichten,
  and Hill}}]{hill}
\bibinfo{author}{\bibfnamefont{W.~A.} \bibnamefont{Bardeen}},
  \bibinfo{author}{\bibfnamefont{E.~J.} \bibnamefont{Eichten}},
  \bibnamefont{and} \bibinfo{author}{\bibfnamefont{C.~T.} \bibnamefont{Hill}},
  \bibinfo{journal}{Phys. Rev.} \textbf{\bibinfo{volume}{D68}},
  \bibinfo{pages}{054024} (\bibinfo{year}{2003}), \eprint{hep-ph/0305049}.

\bibitem[{\citenamefont{Bardeen and Hill}(1994)}]{hill1}
\bibinfo{author}{\bibfnamefont{W.~A.} \bibnamefont{Bardeen}} \bibnamefont{and}
  \bibinfo{author}{\bibfnamefont{C.~T.} \bibnamefont{Hill}},
  \bibinfo{journal}{Phys. Rev.} \textbf{\bibinfo{volume}{D49}},
  \bibinfo{pages}{409} (\bibinfo{year}{1994}), \eprint{hep-ph/9304265}.

\bibitem[{\citenamefont{Bardeen et~al.}()\citenamefont{Bardeen, Eichten, and
  Hill}}]{Bill}
\bibinfo{author}{\bibfnamefont{W.~A.} \bibnamefont{Bardeen}},
  \bibinfo{author}{\bibfnamefont{E.~J.} \bibnamefont{Eichten}},
  \bibnamefont{and} \bibinfo{author}{\bibfnamefont{C.~T.} \bibnamefont{Hill}},
  \emph{\bibinfo{title}{Private communication}}.

\bibitem[{\citenamefont{Mattson}(2002)}]{Thesis}
\bibinfo{author}{\bibfnamefont{M.~E.} \bibnamefont{Mattson}}
  (\bibinfo{collaboration}{SELEX}) (\bibinfo{year}{2002}),
  \eprint{FERMILAB-THESIS-2002-03}.

\bibitem[{\citenamefont{Russ et~al.}(1998)}]{SELEX}
\bibinfo{author}{\bibfnamefont{J.}~\bibnamefont{Russ}} \bibnamefont{et~al.}
  (\bibinfo{collaboration}{SELEX}) (\bibinfo{year}{1998}),
  \eprint{hep-ex/9812031}.

\bibitem[{\citenamefont{Engelfried et~al.}(1999)}]{RICH}
\bibinfo{author}{\bibfnamefont{J.}~\bibnamefont{Engelfried}}
  \bibnamefont{et~al.} (\bibinfo{collaboration}{SELEX}),
  \bibinfo{journal}{Nucl. Instrum. Meth.} \textbf{\bibinfo{volume}{A433}},
  \bibinfo{pages}{149} (\bibinfo{year}{1999}).

\bibitem[{\citenamefont{Kaya et~al.}(2003)}]{prod}
\bibinfo{author}{\bibfnamefont{M.}~\bibnamefont{Kaya}} \bibnamefont{et~al.}
  (\bibinfo{collaboration}{SELEX}), \bibinfo{journal}{Phys. Lett.}
  \textbf{\bibinfo{volume}{B558}}, \bibinfo{pages}{34} (\bibinfo{year}{2003}),
  \eprint{hep-ex/0302039}.

\bibitem[{\citenamefont{Iori et~al.}(2001)}]{lifetime}
\bibinfo{author}{\bibfnamefont{M.}~\bibnamefont{Iori}} \bibnamefont{et~al.}
  (\bibinfo{collaboration}{SELEX}), \bibinfo{journal}{Phys. Lett.}
  \textbf{\bibinfo{volume}{B523}}, \bibinfo{pages}{22} (\bibinfo{year}{2001}),
  \eprint{hep-ex/0106005}.

\bibitem[{\citenamefont{Balatz et~al.}(2004)}]{TM}
\bibinfo{author}{\bibfnamefont{M.}~\bibnamefont{Balatz}} \bibnamefont{et~al.}
  (\bibinfo{collaboration}{SELEX}), \emph{\bibinfo{title}{The lead-glass
  calorimeter for the selex experiment}} (\bibinfo{year}{2004}),
  \eprint{Fermilab/TM/2252(2004)}.

\bibitem[{\citenamefont{Hagiwara et~al.}(2002)}]{PDG}
\bibinfo{author}{\bibfnamefont{K.}~\bibnamefont{Hagiwara}} \bibnamefont{et~al.}
  (\bibinfo{collaboration}{PDG2002}), \bibinfo{journal}{Phys. Rev.}
  \textbf{\bibinfo{volume}{D66}}, \bibinfo{pages}{010001}
  (\bibinfo{year}{2002}).

\bibitem[{\citenamefont{Kushnirenko et~al.}(2001)}]{d0}
\bibinfo{author}{\bibfnamefont{A.}~\bibnamefont{Kushnirenko}}
  \bibnamefont{et~al.} (\bibinfo{collaboration}{SELEX}),
  \bibinfo{journal}{Phys. Rev. Lett.} \textbf{\bibinfo{volume}{86}},
  \bibinfo{pages}{5243} (\bibinfo{year}{2001}), \eprint{hep-ex/0010014}.

\bibitem[{\citenamefont{Cronin-Hennessy and Galik}()}]{Galik}
\bibinfo{author}{\bibfnamefont{D.}~\bibnamefont{Cronin-Hennessy}}
  \bibnamefont{and} \bibinfo{author}{\bibfnamefont{R.}~\bibnamefont{Galik}}
  (\bibinfo{collaboration}{CLEO}), \emph{\bibinfo{title}{Private
  communication}}.

\bibitem[{\citenamefont{Aubert et~al.}(2004)}]{Babar2}
\bibinfo{author}{\bibfnamefont{B.}~\bibnamefont{Aubert}} \bibnamefont{et~al.}
  (\bibinfo{collaboration}{BABAR}) (\bibinfo{year}{2004}),
  \eprint{hep-ex/0408087}.

\bibitem[{\citenamefont{Kutschke}()}]{Focus}
\bibinfo{author}{\bibfnamefont{R.}~\bibnamefont{Kutschke}}
  (\bibinfo{collaboration}{FOCUS}), \emph{\bibinfo{title}{Private
  communication}}.

\end{thebibliography}

\end{document}